\documentclass[aps,prb,preprint,superscriptaddress,showpacs]{revtex4}

\usepackage{graphicx}
\usepackage{dcolumn}
\usepackage{bm}

\begin{document}

\title{Stabilization of charge and orbital ordered states by antiferromagnetism in the A-site ordered YBaMn$_2$O$_6$}

\author{X. N. Zhang}

\affiliation{Stanford Synchrotron Radiation Laboratory, 
Menlo Park, CA 94025, USA}

\affiliation{Max Planck Institute for Solid State Research, D-70569
Stuttgart, Germany}

\author{K.-Y. Choi}

\affiliation{Department of Physics, Chung-Ang University, 
Seoul 156-756, Republic of Korea}

\author{P. Lemmens}

\affiliation{Max Planck Institute for Solid State Research, D-70569
Stuttgart, Germany}

\affiliation{Institute for Condensed Matter Physics, TU
Braunschweig, D-38106 Braunschweig, Germany}

\author{T. Nakajima}

\affiliation{Nat. Institute of Advanced Industrial Science and Technology, Tsukuba, Ibaraki
305-8565, Japan}

\author{Y. Ueda}

\affiliation{Institute for Solid State Physics, University of Tokyo, 5-1-5 Kashiwanoha, Kashiwa, Chiba 277-8581, Japan}

\date{\today}

\begin{abstract}
We report on Raman scattering measurements of the A-site ordered YBaMn$_2$O$_6$ in the wide temperature range of $10-650$~K. Upon cooling almost all phonon modes show pronounced temperature dependencies of frequency, linewidth, and scattering intensity. This suggests that the octahedra tilting degrees, Mn-O bond angles and lengths vary with temperature in a concerted way. The remarkable feature is that the high-temperature metallic and the charge and orbital ordered (CO-OO) states coexist down to $T_{N}$=195~K. In the antiferromagnetic state zone-folded modes are induced by the CO-OO superstructure and the phonon modes undergo an anomalous softening and narrowing.  Our results evidence that the CO-OO state is stabilized by the spin ordering for temperatures below $T_{N}$.

\end{abstract}


\maketitle


Over the last decade the manganese perovskites R$_{1-x}$A$_x$MnO$_3$ (R=rare earth, A=alkaline earth) have
attracted intense theoretical and experimental attention in condensed matter physics because of the discovery of colossal magneto-resistance and intrinsic electronic phase separation.~\cite{Dagotto2,Dagotto,Tokura} The rich complexity based on structural, electronic, magnetic, and orbital degrees of freedom leads to a vast variety of phases such as paramagnetic insulating, ferromagnetic metallic and charge-ordered antiferromagnetic phases as a function of doping and/or temperature. The stability of these phases relies strongly on the A-site ionic size (the tolerance factor $f$) and randomness.

In contrast to the ionic size of the A-site cations, the degree of disorder can hardly be controlled at finite $x$ because R$^{3+}$ and A$^{2+}$ ions are randomly distributed in the lattice. This puts an obstacle to the understanding of the role of the A-site order/disorder for the complex phase diagram. Recently, a controlled occupation of the R$^{3+}$ and A$^{2+}$ ions has been achieved in the case of half-substituted systems R$_{1/2}$A$_{1/2}$MnO$_3$, which can be successfully synthesized as RBaMn$_2$O$_6$ (R=Y,Tb,Sm).~\cite{Nakajima02,Akahoshi03,Kageyama,Nakajima04,Akahoshi04,Nakajima05,Williams05,Williams05b,Kawasaki06,Aladine,Kawasaki09} This new class of metal-ordered double perovskites enables us to compare in detail ordered with disordered manganites.

The RBaMn$_2$O$_6$ compounds have a tetragonal structure, P4/mmm. The MnO$_2$ square layers are stacked alternately by RO and BaO layers along the $c$ axis. The big difference of the ionic radii between Ba$^{2+}$ and R$^{3+}$ introduces novel structural aspects; (i) strongly distorted MnO$_6$ octahedra and (ii) an their anomalous tilting. This is contrasted by the disordered R$_{0.5}$Ba$_{0.5}$MnO$_3$ which has a primitive cubic structure without a tilting of the MnO$_6$ octahedra. The disordered compound has a spin glassy ground state and semiconducting behavior.~\cite{Akahoshi03} In contrast, the ordered compounds exhibit a transition from a metallic to a charge-ordered (CO) state at high temperatures and an antiferromagnetic ground state. This suggests that the absence of the A-site disorder stabilizes the CO phase while suppressing magneto-resistance effects.~\cite{Nakajima05}

Hereafter, we focus our attention on YBaMn$_2$O$_6$, which shows the highest orbital, charge, and antiferromagnetic ordering temperatures among RBaMn$_2$O$_6$.~\cite{Kageyama} It undergoes three successive transitions. First,  a first-order structural phase transition takes place from a pseudo-orthorhombic to a pseudo-tetragonal state at $T_{t}=520$~K from the paramagnetic metallic states (PM$_1$ to PM$_2$). Second, a metal-insulator transition (MIT) occurs at $T_{CO}=480$~K from the PM$_2$ to paramagnetic insulating (PI) state. Third, the PI state transits to an antiferromagnetic ordered (AFI) state at $T_{N}=195$~K. Two different models have been proposed for the concomitant charge and orbital ordering (CO-OO). In the ionic model the simple Mn$^{3+}$/Mn$^{4+}$ rock-salt-type order is accompanied by the ferro-orbital ordering in which d$_{z^2}$-type Mn$^{3+}$ ions are arranged in a parallel fashion.~\cite{Williams05} The alternative Zener polaron ordering picture supports a formation of ferromagnetic plaquettes consisting of four Mn atoms between $T_N<T<T_{CO}$, which evolves into a noncollinearly ordered state.~\cite{Aladine}

In this brief report we investigate charge and lattice dynamics by Raman spectroscopy, which is sensitive to orbital order and structural changes via optical phonons. We find the coexistence of the metallic and CO-OO states over a wide temperature range and the stabilization of the new CO-OO pattern by spin ordering.

Polycrystalline samples of YBaMn$_2$O$_6$ were prepared by a conventional solid state reaction. Raman scattering
measurements were performed in a quasi-backscattering geometry with the excitation line $\lambda= 514.5$ nm of an Ar$^{+}$ laser. The incident power of 10 mW avoids significant heating and irradiation effects. Raman spectra were collected by a DILOR-XY triple spectrometer and a nitrogen cooled charge-coupled device detector. The high temperature measurements from 300~K to 650~K were carried out using a heating stage under vacuum.

\begin{figure}[tbp]
\linespread{1}
\par
\includegraphics[width=11cm]{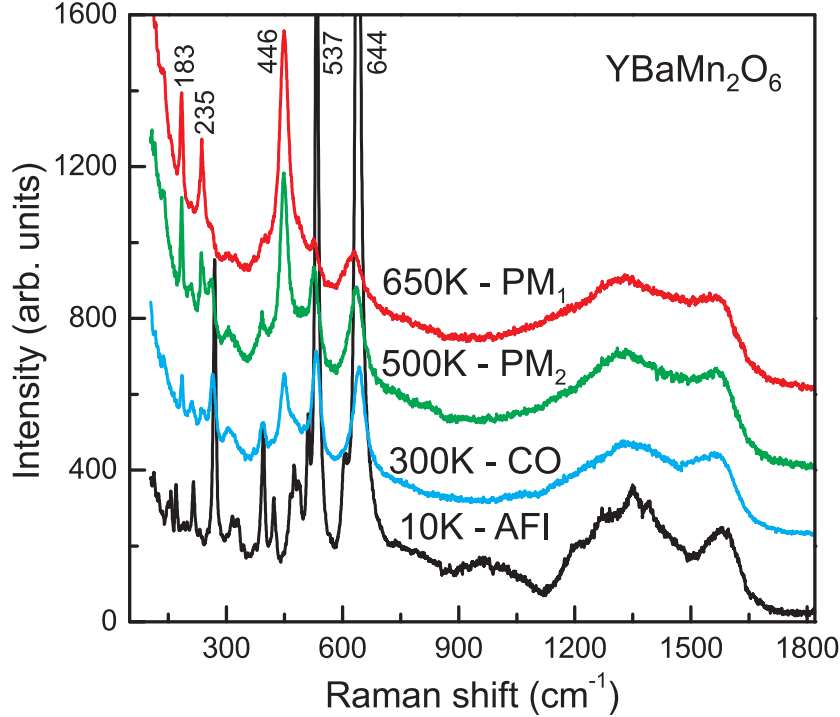}
\par
\caption{Representative Raman spectra of YBaMn$_2$O$_6$
         in four different ordered states: the paramagnetic metallic state (PM$_1$)
          at T=650~K, the paramagnetic metallic state (PM$_2$) at 500~K, the paramagnetic
           insulating state (PI) at 300~K, and the antiferromagnetic (AFI) ordered state at 10~K.
          The spectra are vertically shifted for clarity.} \label{fig:1}
\end{figure}

Figure~1 displays representative Raman spectra of YBaMn$_2$O$_6$ in four different states. In the high-temperature paramagnetic state we resolve 11 phonon modes in the frequency range up to 700~cm$^{-1}$. In spite of metallic conductivity, the 183-, 235-, and 446-cm$^{-1}$ modes are intense and sharp. This is due to the strong distortions of the MnO$_6$ octahedra including tilting. We note that although the real lattice symmetry is lower than the cubic perovskite structure, the main phonon bands can be largely classified in terms of the three cubic groups.~\cite{Choi05} The low-energy external modes of 183 and 235~cm$^{-1}$ are characterized by rotational vibrations of the MnO$_6$  octahedra and are sensitive to the tilting of the MnO$_6$ octahedra. The mid-frequency bending mode of 446-cm$^{-1}$ is susceptible to a change in the Mn-O bond angles. The high-energy stretching modes of 526 and 640~cm$^{-1}$ are assigned to the Jahn-Teller and breathing modes, respectively. Both modes are weak due to the itinerant character of Mn$^{3+}$ related states. With decreasing temperature through the CO state both modes become dominant. This roughly agrees with an earlier study of SmBaMn$_2$O$_6$ that focussed on the Jahn-Teller mode in the transition regime.\cite{Akahoshi04} More significantly, the Raman spectra resemble those of the CE-type state of the layered manganite La$_{1/2}$Sr$_{3/2}$MnO$_{4}$.~\cite{Yamamoto98} This indicates that the CO-OO accompanies both Jahn-Teller and breathing mode type lattice distortions. In addition, the CO-OO activates new phonon modes. As a consequence, we can resolve 24 phonon modes at T=10~K.

We turn next to the pronounced multiphonon scattering in the frequency range of $850 - 1700$~cm$\rm ^{-1}$.
The higher-order bands correspond to overtones and combinations of the 537 and 644~cm$^{-1}$ modes. Since the multiphonon modes are present even in the metallic state, they are not caused by orbital ordering. This was discussed as the origin of the multiphonon scattering in LaMnO$_3$.~\cite{Choi05} With decreasing temperature below the AFI state, the higher-order phonon bands are better resolved with several fine structures. This reflects the formation of a superstructure related to CO-OO. In addition to the contribution of the orbital dynamics we have also to consider resonance Raman scattering. To differentiate between the latter two contributions more detailed information on the electronic band structure would be needed.

\begin{figure}[tbp]
\linespread{1}
\par
\includegraphics[width=11cm]{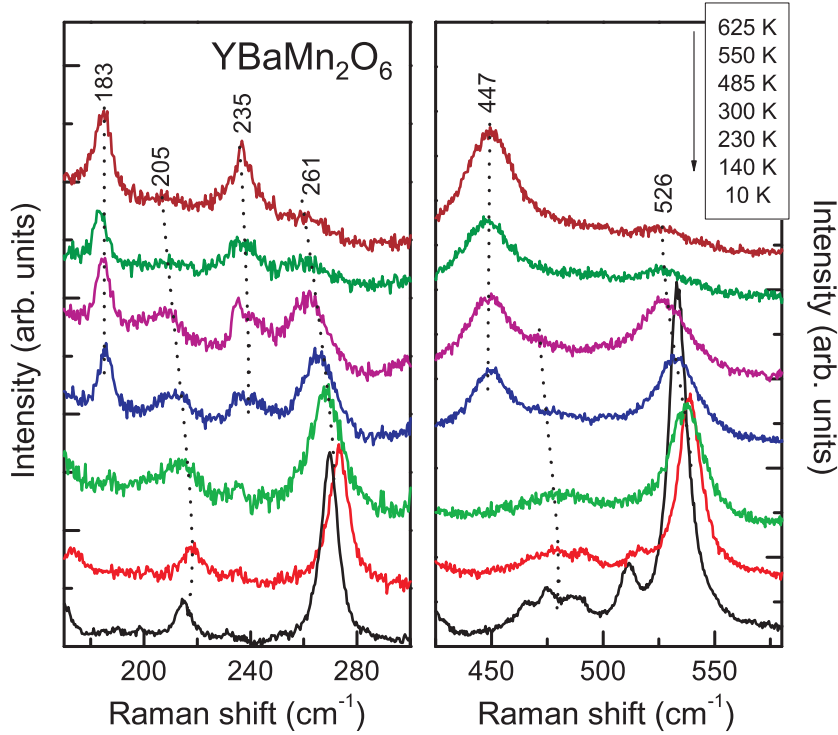}
\par
\caption{The detailed temperature dependence of Raman spectra
          of YBaMn$_2$O$_6$ at low (left) and high (right) energies.
          The spectra are vertically shifted for clarity.}  \label{fig.2}
\end{figure}

In Fig.~2 we give a more detailed temperature dependence zooming into the low- (left panel) and the high-frequency (right panel) part Raman spectral range. In the PM state there exist weak but detectable peaks at 205 and 261~cm$^{-1}$ on the high energy side of the major rotational modes at 183 and 235~cm$^{-1}$. The 183- and 235-cm$^{-1}$ modes disappear at about $T_{N}$=195~K while the 205- and 261-cm$^{-1}$ modes grow gradually in intensity upon cooling and exhibit a maximum in intensity at the lowest measured temperature. Also, the 447-cm$^{-1}$ bending mode is suppressed below $T_{N}$ while the 475-cm$^{-1}$ mode on the high energy shoulder develops a three-peak structure. This points to the formation of a superstructure. The 526-cm$^{-1}$ Jahn-Teller mode is barely detectable in the metallic state. In the AFI state it becomes intense while new modes are split from the major peak. This is the zone-folded mode caused by the charge ordering. The 644-cm$^{-1}$ breathing mode exhibits a similar behavior (not shown here).

We likely relate the pairwise modes to two inequivalent MnO$_6$ octahedra.~\cite{Nakajima04} The Raman scattering intensity is proportional to the partial derivative of the dielectric function with respect to the amplitude of the normal mode. It is therefore hard to understand why the scattering intensity of the normal modes of the two inequivalent MnO$_6$ octahedra is drastically different. Rather, we assign the former to the high-temperature metallic state while the latter to the CO state. This is supported by the fact that the MnO$_6$ octahedral tilting increases from two- to three-tilt upon cooling through $T_t$.~\cite{Nakajima04} The Mn-O bond lengths tend to shorten while the bond angles have a tendency to increase. This explains the larger frequency of the low-temperature modes than those of the high-temperature ones as well as the continuous increase of the scattering intensity of the low-temperature modes.

\begin{figure}[tbp]
\linespread{1}
\par
\includegraphics[width=8cm]{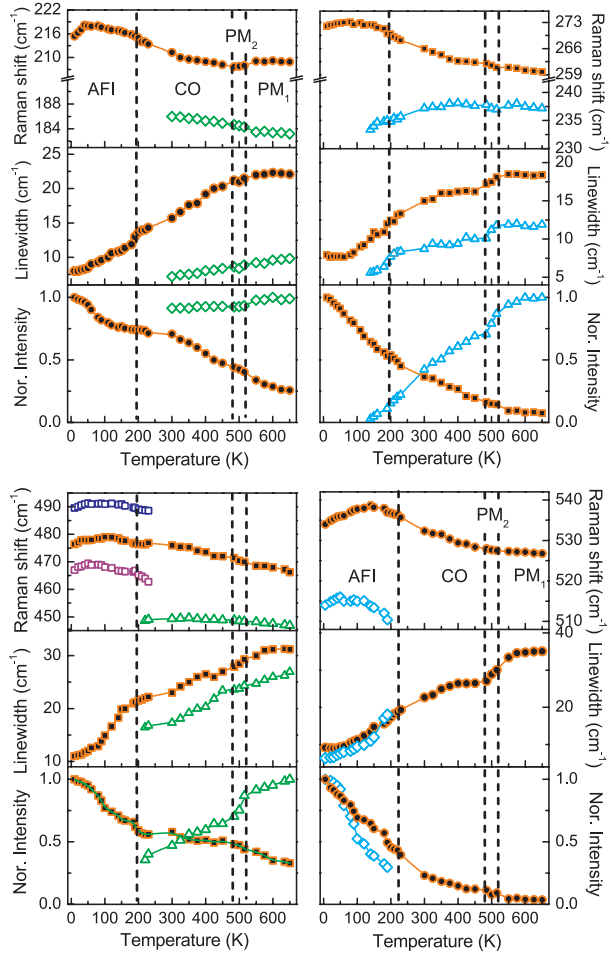}
\par
\caption{ Temperature dependence of Raman shift, linewidth, and
normalized scattering intensity of the 183, 205, 235, 261, 447, 468, and 526-cm$^{-1}$ modes.}  \label{fig.3}
\end{figure}

To analyze the anomalous features of the phonon modes quantitatively we fit them to Lorentzian profiles. The resulting frequency, linewidth, and intensity of the representative modes are summarized in Fig.~3. Here we note that for technical reasons we did not measure the spectra in the temperature range $230 - 290$~K. However, this does not establish any obstacle in extracting the main physics because the respective temperature range corresponds to only to the CO state where all phonon parameters change smoothly (see below).

The phonon modes show a similar distinct temperature dependence. With decreasing temperature the 205- and 261-cm$^{-1}$ rotational modes (full circles and squares) undergo a substantial hardening by 10~cm$^{-1}$ down to 50~K and then
exhibit a tiny softening. Their linewidth  decreases monotonically with an indication for a saturation at temperatures below about 50~K, well below the AFI state. The scattering intensity increases  continuously with a tendency for saturation in a similar temperature range. The 183~cm$^{-1}$ mode (open diamonds), which is the counterpart of the 205~cm$^{-1}$ one, hardens and narrows upon cooling. It disappears abruptly in the vicinity of the magnetic ordering at $T_{N}$=195~K. The 235~cm$^{-1}$ mode (open triangle), which is the counterpart of the 261~cm$^{-1}$ one, decreases quasi-linearly in intensity and is hardly seen below 140~K. With lowering temperature its frequency is shifted down by 4~cm$^{-1}$ and its linewidth decreases with discernible anomalies at the phase transition temperatures. We note that in the A-site disordered manganites the low-frequency modes show no appreciable change.~\cite{choi03}

With decreasing temperature the 468-cm$^{-1}$  bending  mode hardens by 12~cm$^{-1}$ and then a slight softening follows towards lower temperatures. The linewidth and intensity exhibit a kink at about $T_{N}=195$~K. Remarkably, the zone-folded modes are resolved just above the AFI state. The counterpart 468-cm$^{-1}$ mode is gradually suppressed at $T_{N}$. The 526-cm$^{-1}$ Jahn-Teller mode becomes pronounced upon cooling and splits just below $T_{N}$. This indicates that the Jahn-Teller active Mn$^{3+}$ ions grow in number. Overall, the phonon parameters show weak anomalies at the phase transitions. In stead, the phonon anomalies evolve gradually over almost the whole temperature range and all phonon modes exhibit a similar behavior. This points to a concomitant variation of the octahedra tilting degree, Mn-O bond angles and lengths with temperature as well as to a continuous evolution of the CO-OO state even below $T_{CO}$. Furthermore, we note that the scattering intensities behave like the resistivity [Compare to Fig.~5 of  Ref.~\cite{Nakajima04}] and the temperature dependence of the linewidths resembles that of the g-factor above $T_N$ [Compare to Fig.~3 of Ref.~\cite{Zakharov}]. Since the g-factor scales with the magnetic susceptibility in the studied compounds, we conclude that the correlation between the linewidths and the g-factor evidences strong coupling lattice dynamics and spin degrees of freedom.

In the following we will discuss the implication of our results with respect to the CO-OO state. In the high-temperature metallic state above $T_{t}$ a structural study shows that all Mn lattice sites are equivalent.~\cite{Williams05b} Furthermore, ESR measurements provide evidence for a highly mobile polaronic hopping of the $e_g$-electron system.~\cite{Zakharov} In spite of a screening of the phonon modes by the conducting electrons, the Raman spectra are characterized by sharp peaks. This signifies heavy distortions of the MnO$_6$ octahedron. The more salient feature is that the main peaks of the metallic high temperature phase are accompanied by weak peaks, which develop into strong peaks at lower temperatures. This suggests that a small fraction of a local charge ordered state is still present in the background of the metallic state. Also in a neutron scattering study a two phase region has been observed between $T_{t}=520$~K and $T_{CO}=480$~K.~cite{Williams05b} This is further supported by anomalies of the g factor in ESR experiments, that indicate the presence of clusters of the high-temperature phase.~\cite{Zakharov} Therefore we suggest that the phase separation is due to an intrinsic phenomenon and not due to impurities or inhomogeneities.

In the CO state ESR shows a slowing down of the polaronic hopping processes and the persistence of charge fluctuations down to $T=410$~K.~\cite{Zakharov} For $T_N<T<T_{CO}$ our Raman spectra give no hint for the presence of CO induced phonons. Instead, the coexistence of metallic and CO states are observed similarly to observations in the metal-disordered Nd$_{1/2}$Sr$_{1/2}$MnO$_3$.~\cite{choi03}  The retainment of the metallic state in the CO state might be due to a kinetic energy gain of the e$_g$ electrons. Two conflicting models have been proposed to describe the CO-OO state~\cite{Williams05,Aladine}: (i) a Mn$^{3+}$/Mn$^{4+}$ rock-salt-type charge order with a ferro-orbital ordering of d$_{z^2}$-type Mn$^{3+}$ ions, or (ii) a Zener polaron ordering of four Mn ferromagnetic plaquettes. Our method is not able to pin down the exact CO-OO pattern. However, we can say that the CO-OO state between $T_N$ and $T_{CO}$ is more complex than the two models.

Upon cooling through $T_N$, a rearrangement of the orbital stacking pattern takes place.~\cite{Kageyama} In the AFI phase zone-folded modes related to a superstructure appear while phonon modes related the metallic state are suppressed. At the same time, the phonon modes undergo a substantial softening, a narrowing of their linewidth and drastic increase of their scattering intensity. This indicates that new CO-OO patterns are stabilized by the spin ordering. The related strong spin-lattice coupling is confirmed by the observed substantial softening, which is opposite to the expected hardening due to anharmonic lattice interactions.


To conclude, our Raman scattering study of the A-site ordered YBaMn$_2$O$_6$ show pronounced phonon anomalies based on continuous and concomitant changes of the MnO$_6$ octahedra tilting degrees, bond angles and lengths. We find evidence that in this compound metallic and CO-OO states coexist between $T_{N}$ and $T_{CO}$. Furthermore new CO-OO patterns are stabilized below $T_{N}$.

Acknowledgement: We acknowledge important discussion with J. Deisenhofer. This work was partially supported by the DFG and the ESF program \emph{Highly Frustrated Magnetism}. KYC acknowledges financial support from the Alexander-von-Humboldt Foundation.

\end{document}